# Influence of gas ambient on charge writing at the LaAlO$_3$/SrTiO$_3$ heterointerface

*Haeri Kim,[†] Seon Young Moon,[‡] Shin-Ik Kim,[‡,§] Seung-Hyub Baek,[‡,§] Ho Won Jang,[∥] and Dong-Wook Kim*[†]*

[†] Department of Physics, Ewha Womans University, Seoul 120-750, Korea

[‡] Electronic Materials Research Center, Korea Institute of Science and Technology (KIST), Seoul 136-791, Korea

[§] Department of Nanomaterials Science and Technology, University of Science and Technology, Daejeon, 305-333, Republic of Korea

[∥] Department of Materials Science and Engineering, Research Institute of Advanced Materials, Seoul National University, Seoul 151-742, Korea

KEYWORDS: LaAlO$_3$/SrTiO$_3$, Carrier density, Ambient effect, Charge writing

We investigated the influences charge writing on the surface work function and resistance of the LaAlO$_3$/SrTiO$_3$ (LAO/STO) heterointerface in several gas environments (air, O$_2$, N$_2$, and H$_2$/N$_2$). Charge writing decreased the surface work function and resistance of the LAO/STO sample quite a lot in air but slightly in O$_2$. The interface carrier density was extracted from the measured sheet resistance and compared with that obtained from the proposed charge-writing mechanisms, such as carrier transfer via surface adsorbates and surface redox. Such quantitative analyses suggested that additional processes (e.g., electronic state modification and electrochemical surface reaction) were required to explain charge writing on the LAO/STO interface.

## 1. INTRODUCTION

High-mobility, two-dimensional (2D) conduction behaviors have been observed at the LaAlO$_3$/SrTiO$_3$ (LAO/STO) heterointerface.[1-24] Because both materials are wide-bandgap insulators, intensive experimental and theoretical researches have been continued to explain such extraordinary transport behavior.[1] The LAO/STO heterointerface has presented numerous unique physical properties, including a metal–insulator transition,[2–9] superconductivity,[10] magnetic ordering,[11–13] thermoelectricity,[14] electron correlation,[15] non-volatile conductance control,[16] and a huge photoresponse.[17] The ongoing discovery of such phenomena has stimulated much research activity in relevant communities.

Among the variety of fascinating phenomena, modulation of local conductance using biased tip scanning (i.e., 'charge writing') has received a great deal of attention from device applications as well as academic studies. It has been a long-standing obstacle to realize nanoscale patterning of metal oxides.[24–27] Etching damage caused by chemical and/or physical attack often severely deteriorates the physical properties of metal oxides. Thus, it is rather difficult to find in-depth studies on size effects and quantum transport in metal oxides. Recently, charge writing has been successfully used to fabricate oxide-based nanoelectronic devices.[2–9] Surface adsorption and field-induced local desorption of water molecules (the water-cycle mechanism) have been suggested as the origin of conductance modification.[7,18] Nanoscopic redox and the resulting increase in the carrier concentration offer another possibility.[19] Although experimental results led researchers to propose these scenarios, the mechanism has not been clarified to date.[6–9,18–21]

In this work, we investigated the influence of ambient gas and charge writing on the surface work function and resistance of the LAO/STO heterointerface in four different gases (air, $O_2$, $N_2$, and $H_2/N_2$). The charge writing experiments decreased the work function and resistance of the sample, revealing significant gas ambient dependence. First, we tried to explain such ambient dependence in terms of carrier transfer via charged surface adsorbates and oxygen vacancies. The change in the carrier density was calculated from the difference in the work function after charge writing; this value was then compared with that estimated from the measured conductance. The comparison showed that the charge writing could not be well explained by the removal of charged adsorbates or the formation of oxygen vacancies. These results suggested that additional processes, including electronic state modification and electrochemical reaction, should be considered together to explain the charge writing phenomena.

## 2. EXPERIMENTAL SECTION

**2.1. Sample preparation.** LAO layers, with four unit cells, were deposited by pulsed laser deposition (PLD) on $TiO_2$-terminated (001) STO substrates. A KrF excimer laser beam (wavelength: 248 nm; energy density: 1.5 J cm$^{-2}$; repetition rate: 2 Hz) was focused onto LAO single-crystal targets. The substrates were attached to a resistive heater and positioned 45–50 mm from the target. The LAO films were grown at a substrate temperature of 700°C in an oxygen pressure of 1 mTorr. Detailed growth conditions and structural characterizations can be found in the authors' earlier publication.[16]

**2.2. Resistance measurement.** The resistance of the LAO/STO heterointerface sample was measured using the four-point measurement method. The conducting channel had a square shape, with an area of 10×10 μm$^2$, as shown in Figures 1a and 1b. The conducting region was confined only in the desired area by patterning an amorphous $Al_2O_3$ layer onto the STO substrates, as shown in the cross-sectional schematic diagram of Figure 1c.[24] An amorphous LAO (a-LAO) layer formed on the $Al_2O_3$ layer. An epitaxial LAO (e-LAO) layer formed on the STO substrate; note that e-LAO growth occurred only on the STO substrate. The difference in the refractive indices and configuration of the top layers allowed the patterned area to be easily identified using an optical microscope, as shown in Figure 1b.

**2.3. Work function measurement.** The surface work function of the sample was measured by Kelvin probe force microscopy (KPFM) using an atomic force microscopy (AFM) system (XE-100, Park Systems Co.) with a glove box. Four different ambient gases, air (relative humidity: 20–30%), $O_2$, $N_2$, and $H_2(2\%)/N_2(98\%)$, were contained within a glove box for the measurements. Transport and KPFM experiments were simultaneously performed in each of the ambient gas environments. The sample block was equipped with a heater and a printed circuit board (PCB) for electrical connection, as shown in Figure 1d. All of the measurements were conducted in the dark to exclude the possible influence of photocurrent (see Supporting Information Figure S1). Conductive Pt-coated Si cantilevers (NSG10/Pt, resonance frequency: ~240 kHz, NT-MDT) were used for charge writing and imaging of the LAO/STO sample surface. The charge-writing experiment was performed in contact mode by applying +10 V to the tip, which had a scan speed of 150 nm $s^{-1}$. The KPFM images were acquired by applying an alternating-current (AC) modulation voltage of 2 V and a frequency of 20 kHz to the tip, with a scan speed of 400 nm $s^{-1}$.

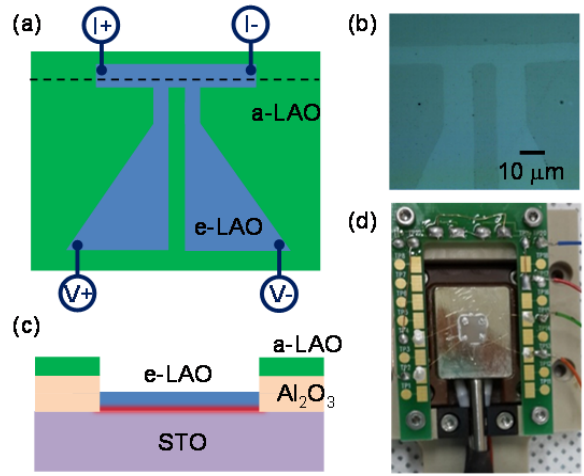

**Figure 1.** (a) Schematic diagrams of the LaAlO$_3$/SrTiO$_3$ (LAO/STO) sample for resistance measurements [green: amorphous-LAO (a-LAO) and blue: epitaxial-LAO (e-LAO)]. (b) Optical microscope image of a patterned LAO/STO sample used in the four-point measurement. (c) Cross-sectional view of the dashed-line in (a). (d) Optical microscopy image of the sample loaded on a heating stage for simultaneous Kelvin probe force microscopy (KPFM) and transport measurements.

## 3. RESULTS AND DISCUSSION

**3.1. Resistance and work function in H$_2$/N$_2$ and O$_2$.** Figure 2a shows the sheet resistance of the LAO/STO sample measured during repeated gas exchange cycles in H$_2$/N$_2$ and O$_2$ (detailed experimental procedures can be found in Supporting Information Figure S2). The average resistance in H$_2$/N$_2$ is slightly smaller than that in O$_2$. However, the variation of the resistance in each gas is very large: the resistance in H$_2$/N$_2$ is sometimes larger than that in O$_2$ (e.g., 1■ > 2●). Thus, the ambient dependence of the resistance is not obvious. As shown in Figure 2b, the surface work function, measured using KPFM, exhibits more clear ambient dependence: all the work function values measured in H$_2$/N$_2$ are smaller than those in O$_2$. This reveals that the surface work function is more sensitive to the ambient than the resistance of the sample.

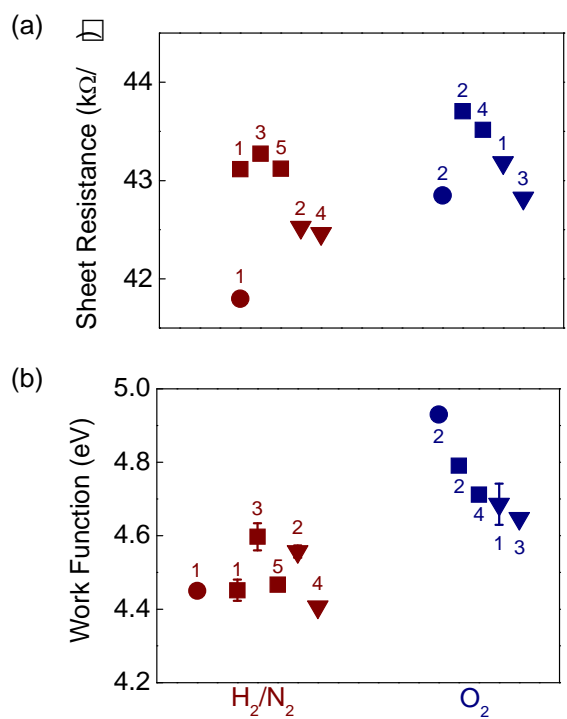

**Figure 2.** (a) Sheet resistance and (b) work function measured in $H_2/N_2$ and $O_2$. The error bars in the work function data indicate standard deviations of the data obtained from 2×2 μm$^2$ area scans. Symbols (●, ■, and ▼) and number represent the data obtained from three different sets of gas exchange cycles and measurement sequence in each set, respectively. Detailed experimental procedures can be found in Supporting Information Figure S2.

Recent theoretical work by Son *et al.* suggested that the adsorption of hydrogen generates surface hydrogen ion ($H^+$) adsorbates and induces electron donation to the LAO/STO interface.[21] The increase in the carrier concentration at the interface would reduce the sample resistance. The existence of charged adsorbates produces surface dipoles and lowers the work function, as illustrated in Figure 3a.[28] Such hydrogen adsorption/carrier donation scenario suggests that the resistance and work function in $H_2/N_2$ should be smaller than those in $O_2$. The work function data (Figure 2b) seem to support the scenario but the variation in the resistance data (Figure 2a) is too large to convince the suggested scenario.

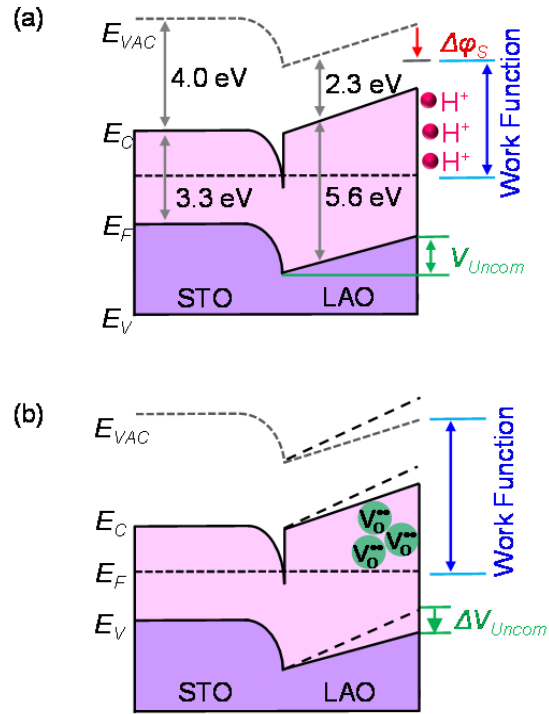

**Figure 3.** Band diagrams illustrating the effects of (a) adsorption and dissociation of $H_2$ to form $H^+$ surface adsorbates and (b) surface redox and lowering of the uncompensated potential, $V_{Uncom}$, due to oxygen vacancies ($V_O^{\bullet\bullet}$).

Most of the complex oxides have predominantly ionic bonds and are prone to a variety of cationic and anionic defects, including vacancies, interstitials, and anti-sites. Defects should be present in the LAO thin film. The creation of a donor state at the surface via a redox reaction and subsequent electron transfer to the interface can increase interface carrier density.[19] At the same time, oxygen vacancies in LAO can decrease the uncompensated potential, $V_{Uncom}$, in the LAO film and the surface work function, as illustrated in Figure 3b. As described in Supporting Information Figure S2, the ambient dependent measurements were repeated while exchanging gas ambient (air → $N_2$ → $H_2/N_2$ → $N_2$ → $O_2$ → $N_2$ → $H_2/N_2$ → $N_2$ → $O_2$ → ⋯ or air → $N_2$ → $O_2$ → $N_2$ → $H_2/N_2$ → $N_2$ → $O_2$ → $N_2$ → $H_2/N_2$ → ⋯). Storage of the sample in $O_2$ would allow dissociation of $O_2$ molecules and the elimination of vacancies in LAO.[22] Thus, the work function and resistance in $O_2$ can be larger than those in

H$_2$/N$_2$. Taken as a whole, the discussion above provides a qualitatively reasonable explanation regarding the ambient effects of the work function.

The hydrogen adsorption scenario allows us to quantitatively estimate the surface coverage from the work function.[28] The potential drop by the dipole is given by $\Delta\varphi_s = e\left(\dfrac{f \times Ned}{\varepsilon}\right)$ ($f$: surface coverage; $N$: number of unit cells per area; $e$: electron charge; $d$: dipole length; and $\varepsilon$: dielectric constant); the potential drop will vary with the surface work function, as shown schematically in Fig. 3(b). The surface coverage of H$^+$, 0.91±0.32%, was estimated from the difference in the work functions in O$_2$ and H$_2$/N$_2$, since negligible amount of H$^+$ adsorbates on the sample surface was expected in O$_2$. Assuming carrier donation from each H atom, the increase in the interface carrier density should be proportional to the surface coverage. If we ignore variations of the mobility in different gases,[23] the increase of interface carrier density in H$_2$/N$_2$ is estimated to be 0.95±0.89% from the measured sheet resistance. These two average values (0.91 and 0.95%) are similar, but the variation in the resistance is too large to well support the hydrogen adsorption and subsequent carrier donation scenario.

**3.2. Ambient effects on charge writing.** Figure 4a and 4b provide a schematic diagram of the charge-writing experiment and the surface work function map of the LAO/STO sample, obtained after charge writing with an AFM tip, respectively. The biased tip was scanned over a local area, having a striped profile (stripe width: 150 nm; stripe length: 10 μm). The resistance drop was measured after writing. The sheet conductance of the area undergoing the charge writing ($G_{stripe}$) can be extracted from the total sheet resistance ($G_{total}$), as shown in Figure 4c. Ambient effects on the charge writing experiments are obvious: the change in the conductance and work function is the largest in air, and the smallest in O$_2$. The decrease in the resistance (−12%) and work function (−0.41 eV) after charge writing in air was similar to

the results reported in the literature.[6–9]

The charge writing can modify the tip due to the application of relatively high voltage to the tip. For estimation of the reliable work function, careful calibration of the tip should be performed just before (or after) every measurement. Such repeated calibration will be really time-consuming. In this work, we measured the relative difference of the work functions obtained from two different areas ('$W_{stripe} - W_{base}$'): charge-writing-experienced region ('stripe') and fresh region ('base'). Such measurement can clearly reveal the influence of the charge writing on the work function in each gas ambient, without any calibration.

Bi *et al*. proposed the water-cycle mechanism as the physical origin of charge writing.[7] In the water-cycle process, water molecules adsorb on the surface and dissociate into $OH^-$ and $H^+$ ions. The negatively charged $OH^-$ ions can be removed by scanning of the positively biased tip. As a result, the surface coverage of $H^+$ can increase and the remaining $H^+$ ions provide carriers to the interface, decreasing the sample resistance. Based on the water-cycle mechanism, the change in the $H^+$ surface coverage can be estimated from the difference in conductance and work function, before and after charge writing. From the conductance increase in air, the $H^+$ surface coverage was estimated to be 881%. In contrast, the coverage extracted from the surface work function in air was only 1.6%. The significant discrepancy raised doubts regarding the validity of the gas adsorption and resulting carrier transfer scenario.

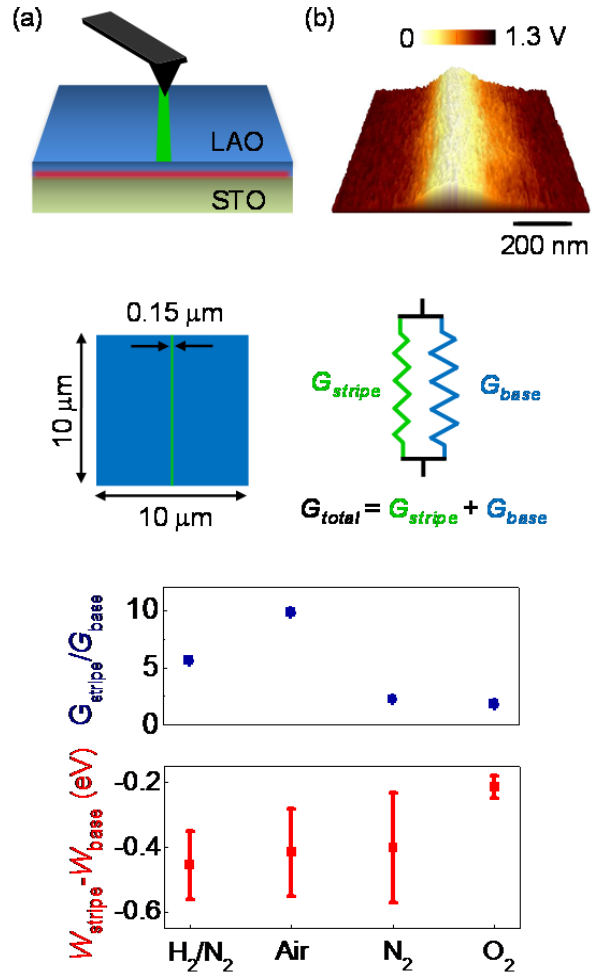

**Figure 4.** (a) Schematic diagram of the positively biased atomic force microscopy (AFM) tip used in the 'charge-writing' process. (b) Surface potential map (1 × 1 $\mu m^2$) after charge writing in a $H_2/N_2$ atmosphere. (c) Schematic diagram of the scanned region ('stripe') and surrounding fresh region ('base'). (d) The sheet conductance ratio ($G_{stripe}/G_{base}$) and the work function difference ($W_{stripe} - W_{base}$) of the scanned and fresh areas.

Field-induced redox was considered as an alternative possibility to explain charge writing. If the positively biased tip can remove oxygen ions from the LAO/STO sample, then oxygen vacancies are generated. Bristowe *et al.* suggested that oxygen vacancies can supply carriers to the interface and lower the uncompensated potential, $V_{Uncom}$.[19] According to such a scenario, the change in the total sheet electron density at the interface ($\Delta\sigma$) can be estimated by the following equation, $\Delta\sigma = -\frac{\varepsilon_{LAO}}{e\, d_{LAO}}\Delta V_{Uncom}$ ($\sigma$: interface carrier concentration, $\varepsilon_{LAO}$: dielectric constant of LAO; and $d_{LAO}$: LAO layer thickness).[23] In $N_2$, the environment has a negligible amount of water and $H_2$; thus, surface redox rather than the water-cycle mechanism may dominantly contribute to the charge writing process. The change in the uncompensated potential, $V_{Uncom}$, is equal to the difference in the work function before and after charge writing (see Figure 3b), since hydrogen adsorption is very unlikely to occur in $N_2$. Therefore, $\Delta\sigma$ can be estimated from the change in the measured surface work function, as well as the amount of conductance increase. In $N_2$, $\Delta\sigma/\sigma$ estimated from the conductance data is 1.3, almost ten times larger than that obtained from the work function change, 0.14. Therefore, field-induced redox (oxygen vacancy generation) alone could not explain the charge writing on the LAO/STO surface, even in hydrogen-free ambient.

Li and Yu reported systematic studies on the dissociation of water molecules in the LAO/STO system, based on first-principles calculations.[20] They argued that the adsorption of water molecules had little effect on the electronic states of the LAO surface; however, the dissociation of water molecules could modify the electronic states of the LAO surface and the LAO/STO interface, leading to a metallic interface. For example, 1/4 surface coverage of water can raise the valence band maximum (VBM) of LAO by ~0.2 eV. If such a process occurs, then the work function of the LAO/STO surface should increase, and the resistance of the interface should decrease. This tendency is opposite to that predicted by the water-cycle mechanism and field-induced redox scenarios; for both of these cases, the work function and the resistance decrease. Thus, this electronic state modification scenario helps to resolve the aforementioned discrepancy about the change of work function and resistance. However, extraction of the carrier density from the measured surface work function variation is not straightforward. The origin of the water molecules should be mentioned: some of the water adsorbates can be removed by the baking process in our experiments (heating at 120°C in $N_2$ gas), but chemisorbed water adsorbates are hard to be removed.[23]

Kumar *et al.* investigated the local electrochemical phenomena of charge writing on the LAO/STO surface, using electrochemical strain microscopy and KPFM.[9] They argued that the charge writing mechanism should include somewhat complicated electrochemical components rather than simple screening by charged surface adsorbates. They also claimed that the electrochemical processes could be attributed to either surface charging due to electrochemical water-layer splitting or oxygen vacancy formation on or close to the material surface. As reported by many researchers, the large electric field produced by the biased tip can cause a variety of physical and electrochemical phenomena at the oxide surface in ambient air (mixture of various gases and water vapor).[28–30] For example, water adsorbates can form pillars of water between the tip and sample surface (capillary condensation), which results in contamination and unwanted electrochemical surface reactions. To avoid such complications, experiments can be performed under high-vacuum conditions.[29] Alternatively, measurements in controlled ambient are helpful to minimize the complicated effects.[28]

Therefore, all the possible processes (i.e., the water-cycle mechanism, field-induced redox, electronic state modification, and electrochemical reaction) should be considered together to understand the charge-writing on the LAO/STO interface. In air, the LAO/STO sample surface is covered with several adsorbates (detailed species can vary, depending on the experimental environment), and hence all the four processes can take place. In dry gases ($H_2/N_2$, $N_2$, and $O_2$), large amount of surface adsorbates can be removed by the baking and residual water adsorbates may induce electronic state modification via dissociation of water molecules. Contribution of other processes may differ depending on specific gas. In $H_2/N_2$, the water-cycle mechanism (by $H_2$ not water vapor) and redox process are more important than others. In $N_2$, redox process will dominate. The redox hardly occurs in $O_2$, since supply of $O_2$ from the ambient will continuously eliminate oxygen vacancies in LAO.[22] In $O_2$ other processes are not very plausible, and hence the dissociative adsorption of the residual water molecules may be important.

Since the discovery of the 2D conduction behaviors at the LAO/STO heterointerface, there has been extensive research activities to explain the exotic transport phenomena: the polar catastrophe mechanism suggests an intrinsic electronic reconstruction and there are also many reports to demonstrate the crucial contribution of defect-driven conductivity.[1] The discussion in our work doesn't require a specific model to explain the origin of the interface carriers. We just assumed that the interface carrier concentration dominantly determined the sheet resistance of the LAO/STO sample, which were supported by experimental results in literature.[23]

## 4. CONCLUSIONS

We investigated the effect of ambient gas on the LAO/STO heterointerface, in particular changes in the surface work function and resistance of the sample. Based on the proposed models, interface carrier concentration could be estimated from the measured work function data, including the dipole contribution of the surface adsorbates and the oxygen-vacancy-induced lowering of the uncompensated potential in the LAO layer. The gas adsorption/dissociation at the surface, followed by charge carrier transfer to the interface, could explain the ambient gas effects on the work function but not well on the resistance. We also studied the charge-writing-induced change in the work function and resistance of the LAO/STO sample in various gas environments. The change in the carrier density, estimated from the measured work function, was much smaller than that given by the measured resistance. This suggested that the scenarios based on the adsorption/desorption of charged adsorbates and surface redox were not sufficient to explain the charge-writing mechanism.

Other processes, including electronic state modification and electrochemical reactions, should be included to explain the tip-induced resistance change of the LAO/STO interface.

## ASSOCIATED CONTENT

**Supporting Information**. Sheet resistance vs. time plot and work flow of the ambient dependent measurements. This material is available free of charge via the Internet at http://pubs.acs.org.

## AUTHOR INFORMATION

**Corresponding Author**

* D.-W. Kim: +82 2 3277 6668; Fax: +82 2 3277 2372; E-mail: dwkim@ewha.ac.kr

**Notes**

The authors declare no competing financial interest.

## ACKNOWLEDGMENT

This work was supported by the Quantum Metamaterials Research Center (2008-0061893) and Basic Science Research Program through the National Research Foundation of Korea Grant, and the Korea Institute of Science and Technology (KIST) through 2E24001. H.W.J. acknowledges the Outstanding Young Researcher Program through the National Research Foundation of Korea.

# Supporting Information

# Influence of gas ambient on charge writing at the LaAlO$_3$/SrTiO$_3$ heterointerface

*Haeri Kim,[†] Seon Young Moon,[‡] Shin-Ik Kim,[‡,§] Seung-Hyub Baek,[‡,§] Ho Won Jang,[∥] and Dong-Wook Kim*[†]*

[†] Department of Physics, Ewha Womans University, Seoul 120-750, Korea

[‡] Electronic Materials Research Center, Korea Institute of Science and Technology (KIST), Seoul 136-791, Korea

[§] Department of Nanomaterials Science and Technology, University of Science and Technology, Daejeon, 305-333, Republic of Korea

[∥] Department of Materials Science and Engineering, Research Institute of Advanced Materials, Seoul National University, Seoul 151-742, Korea

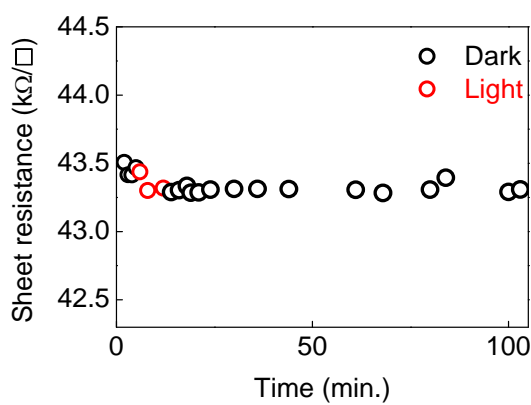

**Figure S1**. Sheet resistance vs. time plot of a LAO/STO sample, obtained during 100 min. including the KPFM measurement time. The sample was stored in dark during the N$_2$ purging and the following baking procedures, as shown in Figure S1. Thus, the first measurement was started after loading the sample in dark after >18 hours. After exposure of a normal fluorescent lamp, we cannot notice clear feature of the photocurrent relaxation after starting the measurement. The time interval denoted by 'Light' in Figure S2 corresponds to the time in which the sample was exposed to the IR (infrared) LED source of the KPFM system. No clear difference can be seen before and after the IR light exposure, except for random fluctuation of the resistance.

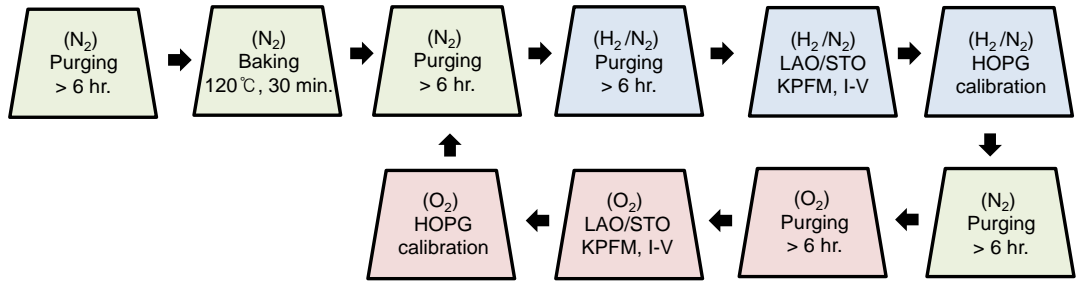

**Figure S2**. Work flow of the ambient dependent KPFM (Kelvin probe force microscopy) and current-voltage (*I-V*) measurements for repeated gas exchange cycles. At first, the glove box was purged by $N_2$ for more than 6 hours. Then, the sample was heated at 120°C for 30 min. while flowing $N_2$ gas using a heater in the sample holder. Prior to changing the gas ambient (from $O_2$ to $H_2/N_2$ or vice versa), the system was purged by $N_2$ for more than 6 hours to remove residual gas and then filled with the intended gas for more than 6 hours. After that, the work function and resistance measurements were done while still flowing the gas to maintain the dynamic equilibrium. Just after each measurement, the tip was calibrated with the HOPG (highly ordered pyrolytic graphite, SPI Supplies) reference sample. A set of measurements take several days. After break of a few days, another experiment was tried again. Different symbol (●, ■, and ▼) in the plot (Figure 2 and 3a) represent the data obtained from each experiment consisting of several gas exchange cycles.

**1st set of experiments)** $N_2$ (baking) → 1st measurement in $H_2/N_2$ [indicated by '1●'] → 2nd measurement in $O_2$ [indicated by '2●']

**2nd)** $N_2$ (baking) → 1st measurement in $H_2/N_2$ [1■] → 2nd measurement in $O_2$ [2■] → 3rd, 4th, and 5th in similar way

**3rd)** $N_2$ (baking) → 1st measurement in $O_2$ [1▼] → 2nd measurement in $H_2/N_2$ [2▼] → 3rd and 4th in similar way